\documentclass{aa}
\usepackage{graphicx}
\usepackage{txfonts}
\usepackage{hyperref}
\begin{document}

\title{Morphokinematical study of the planetary nebula Me\,2-1: Unveiling its point-symmetric and
  unusual physical structure}

   \author{Luis F. Miranda\inst{1}
          \and
          Roberto V\'azquez\inst{2}
          \and
          Lorenzo Olgu\'{\i}n\inst{3}
          \and
          Pedro F. Guill\'en\inst{4}
          \and
          Jos\'e M. Mat\'{\i}as\inst{5}
          %\fnmsep\thanks{Just to show the usage
          %of the elements in the author field}
          }

   \institute{Instituto de Astrofisica de Andalucia (IAA), CSIC, Glorieta de la
     Astronom\'{\i}a s/n, 18008, Granada, Spain\\
     \email{lfm@iaa.es}
     \and
     Instituto de Astronom\'{\i}a, Universidad Nacional Aut\'onoma de M\'exico, Apdo. Postal 877, 22800 Ensenada, B.C.,
     Mexico
     \and
     Departamento de Investigaci\'on en F\'{\i}sica, Universidad de Sonora, Blvd. Rosales Esq. L.D. Colosio, Edif. 3H,
     83190 Hermosillo, Son., Mexico
     \and
     Observatorio Astron\'omico Nacional, Instituto de Astronom\'{\i}a, Universidad Nacional Aut\'onoma de M\'exico,
     Apdo. Postal 106, 22800 Ensenada, B.C., Mexico
     \and
     Departamento de Estad\'{\i}stica e I.O, Escola de Enxe\~ner\'{\i}a Industrial, Universidade de Vigo, Campus As Lagoas-Marcosende, E-36330 Vigo, Spain
             }

   \date{Received ; accepted }

% \abstract{}{}{}{}{} 
% 5 {} token are mandatory
 
   \abstract{Me\,2-1 is a high-excitation planetary nebula whose morphology and physical structure have not yet been investigated. We present narrow-band
     images in several emission lines, and high- and intermediate-resolution long-slit spectra aimed at investigating its morphology and 3D  
     structure, and its physical parameters and chemical abundances.
     By applying deconvolution techniques to the images, we identified in Me\,2-1:  
     an elliptical ring; two elongated, curved structures (caps) that contain three pairs of bright point-symmetric (PS) knots; 
     a shell interior of the ring; and a faint halo or attached shell.
     The caps are observed in all images, while the PS knots are only observed in the low-excitation emission line ones. These structures are also identified in the high-resolution
     long-slit spectra, allowing us to study their morphokinematics. The 3D reconstruction shows that Me\,2-1 consists of a ring seen
     almost pole-on, and a virtually spherical shell, to which the caps and PS knots are attached.
     Caps and PS knots most probably trace the sites where high-velocity collimated bipolar outflows, ejected along a wobbling axis, collide with the spherical shell, 
       are slowed down, and remain attached to it. Although the main excitation mechanism in Me\,2-1 is found to be photoionization, a contribution of shocks
       in the PS knots is suggested by their emission line ratios. The combination of collimated outflows and a ring with a spherical shell is unusual among planetary
     nebulae. We speculate that two planets, each with less than one Jupiter mass, could be involved in the formation of Me\,2-1 if both enter a common
     envelope evolution
     during the asymptotic giant branch phase of the progenitor. One planet is tidally disrupted, forming an accretion disk around the central star, from which collimated bipolar outflows
     are ejected; the other planet survives, causing wobbling of the accretion disk. The physical parameters and chemical
     abundances obtained from our intermediate-resolution spectrum are similar to those obtained in previous analyses, with the abundances also pointing to a low-mass
     progenitor of Me\,2-1. }


   \titlerunning{The point-symmetric planetary nebula Me\,2-1}

   \maketitle
%
%-------------------------------------------------------------------

\section{Introduction}

Me\,2-1 (PN\,G342.1+27.5; $\alpha$(2000.0) = $15^{\rm h}$ $22^{\rm m}$
$19\rlap.^{\rm s}3$; $\delta$(2000.0) = $-23^{\circ}$ 37$'$ 31$''$) is a planetary nebula (PN) discovered by P.\,W. Merril in
an objective-prism plate of April 6, 1940, and described by him as {\it ``a disk about 6\,arcsec in diameter with a conspicuous
  nucleus''} (see Merril 1942). It is a relatively high-excitation PN with a central star (CS) that has an effective temperature,
$T$$_{\rm eff}$, of $\sim$119--230\,kK (Wolff et al. 2000, and references therein). The CS can be seen in the
{\it HST} images shown by Wolff et al. (2000) and Surendiranath et al. (2004, hereafter SPGL04), and its V magnitude is
$\simeq$18.4 (Wolff et al. 2000). To the best
  of our knowledge, stellar absorption or emission lines have not yet been identified in the low- and intermediate-resolution spectra available
  in the literature;
in fact, the CS does not have any assigned spectral type. Recently, Jacoby et al. (2021) present results from {\it Kepler}/K2 that
suggest the possible presence of a companion to the CS with an orbital period of $\simeq$22\,d. The nebular physical properties and chemical
abundances of Me\,2-1 have been
studied by, for example, Aller et al. (1981), Kaler (1985), Moreno et al. (1994), Cuisinier et al. (1996), and SPGL04. In particular, SPGL04 present an analysis
based on {\it IUE} and {\it ISO} spectra combined with the optical spectroscopic data by Aller et al. (1981) and Moreno et al. (1994), covering emission
lines from 1240\,{\AA} to 26\,$\mu$m. From their analysis, SPGL04 conclude that the star progenitor of Me\,2-1 had an initial mass of about 1.5\,M$_{\odot}$ and
a nearly Solar chemical composition.

\begin{table*}
\centering  
\caption{Log of the narrow-band images of Me\,2-1}                           
\begin{tabular}{lcccccc}
  \hline
  \hline
Filter    & $\lambda$$_{\rm c}$   & FWHM       & Exposure time & Seeing   & Spatial resolution \\
          &                      &            &              &          & (deconvolved)  \\  
          &    (\AA)             &  (\AA)     &  (s)      & (arcsec) & (arsec)    \\
\hline

[O\,{\sc iii}]  &  5007  & 30 & 600 &  1.2  & 0.6 \\

[O\,{\sc i}]\tablefootmark{a}    & 6300   & 30 & 900 &  1 &   0.5 \\

H$\alpha$       & 6563   & 9  & 250+300 & 1  &   0.6  \\

[N\,{\sc ii}]   & 6584   & 9  & 600+900 & 1 & 0.6 \\

[S\,{\sc ii}]   & 6731   & 10 & 900 & 1.1 & 0.65\\

\hline
\end{tabular}
\tablefoot{
  \tablefootmark{a}{This filter includes the [S\,{\sc iii}]$\lambda$6312 emission line.}
  }
\end{table*}

\begin{table*}
\centering  
\caption{Log of the long-slit high-resolution MES spectra of Me\,2-1\tablefootmark{a}}
\begin{tabular}{lccccc}
  \hline
  \hline
Date        & Slit PA     & Slit width    & Binning  & Exposure time  & Seeing \\
           &  ($^{\circ}$)  & (arcsec)      &          &  (s)         &  (arcsec) \\  
\hline

2002 July 17 & +90   & 1  & 1$\times$1 & 1800 & 2.6 \\

2004 July 27 & $-$25 & 2 & 2$\times$2 & 1200 & 2.1  \\

             & +40   & 2 & 2$\times$2 & 1200 & 2.1 \\

2004 July 28 & +65N (+1$\farcs$5 N)\tablefootmark{b} & 1 & 1$\times$1 & 1200 & 2.2 \\

             & +65  & 1 & 1$\times$1 & 1200 & 2.1 \\

             & +65S (+1$\farcs$5 S)\tablefootmark{b} & 1 & 1$\times$1 & 1200 & 2.3 \\

%2018 March 21 & +40  & 1 &  4$\times$4 & 900 & 2.5  \\

%               & +90  & 1 &  4$\times$4 & 900 & 3.0  \\

2018 March 21   & $-$15E (+2$\farcs$3 E)\tablefootmark{b}  & 1 &  4$\times$4 & 900 & 2.5 \\

                & $-$35  & 1 &  4$\times$4 & 900 & 2.4 \\
\hline
\end{tabular}
\tablefoot{
  \tablefootmark{a}{The spectra cover the H$\alpha$, [N\,{\sc ii}]$\lambda$$\lambda$6548,6583, and
    He\,{\sc ii}\,$\lambda$6560 emission lines, except that at PA $-$35$^{\circ}$ that covers the
    [O\,{\sc iii}]$\lambda$5007 emission line.}
  \tablefootmark{b}{Displacement and direction of the slit from the center of the nebula.}
  }

\end{table*}

The morphology of Me\,2-1 can be seen in the broadband {\it HST} images and shows a relatively faint, slightly elliptical shell and a brighter
barrel-like structure contained within the elliptical shell (Wolff et al. 2000; SPGL04). Gesicki et al. (1998) claim the existence of a pronounced bipolar
structure in Me\,2-1 on
the basis of the intensity distribution of the [N\,{\sc ii}]$\lambda$6583 emission line in spectra taken at different positions on the nebula. These authors
also mention that the bipolar structure is not observed in the intensity distribution of the H$\alpha$ emission line. Such a pronounced bipolar structure cannot
be recognized in the {\it HST} images, although none of them covers the [N\,{\sc ii}] emission lines. It is obvious that images in, at least, the [N\,{\sc ii}]
and H$\alpha$ emission lines are necessary to clarify the morphology of Me\,2-1. As for its kinematics, the only information available is its
expansion velocities of $\simeq$20 and $\simeq$40\,km\,s$^{-1}$ in H$\alpha$ and [N\,{\sc ii}], respectively (Gesicki et al. 1998). The internal kinematics of Me\,2-1
has not been investigated yet. It is well known that analysis of the morphokinematic (3D) structure of a PN is a key step in identifying the
existing nebular structures, the possible relationships between them, and inferring the mass ejection processes that may have been involved in the formation of these
objects (see, e.g., Akras et al. 2016; Sabin et al. 2017; Miranda et al. 2001, 2017; Danehkar 2022; G\'omez-Mu\~noz et al. 2023). 

In this paper, we present narrow-band optical images in several emission lines, and high- and intermediate-resolution long-slit spectra
of Me\,2-1, with the goal of investigating its morphology and 3D structure, and analyzing its physical and chemical properties. Our
data reveal that Me\,2-1 is a point-symmetric (PS) PN with an unusual combination of structural components. On the basis of the resulting 3D reconstruction,
we discuss how Me\,2-1 may have been formed. 
\begin{figure*}
%\vspace{302pt}
\begin{center}
\includegraphics[width=150mm]{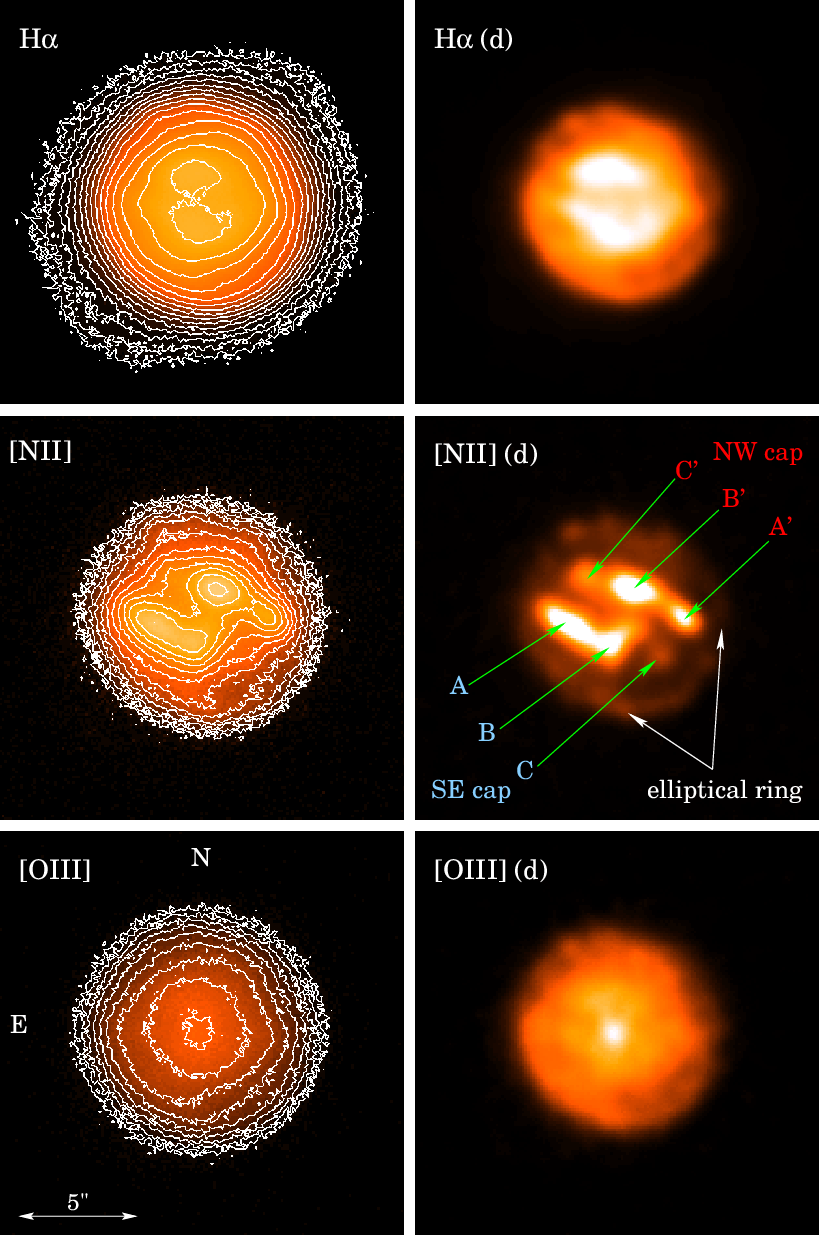}
\caption{Narrow-band H$\alpha$, [N\,{\sc ii}], and [O\,{\sc iii}] images of Me\,2-1. 
({\it Left column}) Original images. Intensity levels and contours are logarithmic. Contours are separated by a factor of 2$^{1/2}$ in linear intensity. ({\it Right column}) Deconvolved images denoted with (d) after the filter's name. Intensity levels
  are linear. The elliptical ring, the SE and NW caps, and the PS knots in the caps (see text) are labeled in the [N\,{\sc ii}](d) image. For the
  labels of the the caps
  and bright knots, blue and red colors are used for blueshifted and redshifted features with respect to the systemic velocity of the nebula. Orientation and spatial scale are identical in the six panels and indicated in the [O\,{\sc iii}] image (bottom left).}
\end{center}
\end{figure*}

\section{Observations}

\subsection{Imaging}

Narrow-band images of Me\,2-1 were obtained on July 24 and 25, 1997 at the Nordic
Optical Telescope (NOT) on Roque de los Muchachos Observatory (ORM, La Palma, Spain). The
detector was a LORAL 2k$\times$2k CCD with a plate scale of
0.11\,arcsec\,pixel$^{-1}$. We used the H$\alpha$ and [N\,{\sc ii}] Instituto de Astrof\'{\i}sica de Canarias (IAC) filters,  
and the [O\,{\sc iii}], [O\,{\sc i}], and [S\,{\sc ii}] NOT filters. The journal of the 
observations is summarized in Table\,1, where we provide the characteristics of the 
filters (central wavelength [$\lambda$$_{\rm c}$] and full width at half maximum (FWHM)), exposure times, and seeing in each image, as indicated by the FWHM of field
stars. The images were cosmic-ray-cleaned, bias-subtracted and flat-fielded using standard routines within the
{\sc midas} package. To gain a better view of the morphology of Me\,2-1, we used the 
Richarson-Lucy algorithm implemented in the {\sc midas} package, and deconvolved the images by using a well-exposed field star
in each image as its point spread function (PSF); the deconvolution process was stopped after 30 iterations when artifacts appeared in the
deconvolved images and affected the image of the nebula. The resulting spatial resolution (FWHM) in the deconvolved images is also provided in Table\,1.

Figure\,1 shows the original and deconvolved (denoted with (d) after the filter's name) H$\alpha$, [N\,{\sc ii}], and [O\,{\sc iii}] images
and Figure\,2 the [O\,{\sc i}](d) and [S\,{\sc ii}](d) images of Me\,2-1. It should be noted that the [O\,{\sc i}]
filter includes the [S\,{\sc iii}]$\lambda$6312 emission line (see Table\,1).

\subsection{High-resolution long-slit spectroscopy}

High-resolution long-slit spectra were obtained with the Manchester Echelle Spectrograph (MES, Meaburn et al. 2003) 
at the 2.1\,m telescope at the San Pedro M\'artir Observatory OAN-SPM. Long-slit spectra were obtained in three epochs: 2002, 2004, and 2018.
In 2002 and 2004 the detector was a SITe3 CCD
that provided spectral and spatial scales of $\simeq$0.051\,{\AA}\,pixel$^{-1}$
and $\simeq$0.3\,arcsec\,pixel$^{-1}$, respectively, in binning 1$\times$1. In 2018, the detector was an E2V CCD that
provides spectral and spatial scales of $\simeq$0.028\,{\AA}\,pixel$^{-1}$
and $\simeq$0.176\,arcsec\,pixel$^{-1}$, respectively, in binning 1$\times$1. In the three epochs, a $\Delta$$\lambda$ = 90\,{\AA} bandwidth filter was used to isolate the
87$^{\rm th}$ order, covering the H$\alpha$, [N\,{\sc ii}]$\lambda$$\lambda$6548,6583, and He\,{\sc ii}\,$\lambda$6560 emission lines.
In 2018, we also used a $\Delta$$\lambda$ = 60\,{\AA} bandwidth filter to
isolate the 114$^{\rm th}$ order, covering the [O\,{\sc iii}]$\lambda$5007 emission line. The long-slit was oriented at several position angles (PAs).
The journal of the MES observations is presented in Table\,2 and the positions of the long-slits are shown in Figure\,3 superimposed on the [N\,{\sc ii}]
image of Me\,2-1. The spectra were reduced using standard procedures within the {\sc iraf} package. After the reduction, the spectra taken in 1$\times$1
binning (Table\,2) were rebinned into a 2$\times$2 binning to increase their signal-to-noise ratio. The achieved spectral resolution  
is $\simeq$12\,km\,s$^{-1}$, as indicated by the FWHM of the ThAr emission lines in the comparison lamp spectra, and the accuracy
of the radial velocity is $\pm$1\,km\,s$^{-1}$. A seeing is included in Table\,2 for each spectrum. 

Position-velocity (PV) maps have been produced from the observed emission lines in all long-slit spectra. Figure\,4 shows PV maps of
the [N\,{\sc ii}]$\lambda$6583 emission line, except at PA +65$^{\circ}$N, which is very similar to that at PA +65$^{\circ}$. Figure\,A1 shows
PV maps of the H$\alpha$ and He\,{\sc ii} emission lines at PAs $-$25$^{\circ}$ and +65$^{\circ}$; the PV maps at other PAs are not shown here because
they are similar to those presented in Fig\,A1. Finally, Figure\,A2 presents the PV map of the [O\,{\sc iii}]$\lambda$5007 emission line at
PA $-$35$^{\circ}$. 

\subsection{Intermediate-resolution long-slit spectroscopy}

Intermediate-resolution long-slit spectra of Me\,2-1 were obtained on August 13, 2004 with the Boller \& Chivens spectrograph mounted
on the 2.1\,m telescope at OAN-SPM. The detector was a SITe3 CCD (24\,$\mu$m\,pix$^{-1}$) with 
a 1k$\times$1k pixel array. We used a 400\,lines\,mm$^{-1}$ dispersion grating along with a 
2$\farcs$5 slit width, giving a spectral resolution of $\sim$6.5\,{\AA} (FWHM). Three exposures of 30\,s each were obtained with the
slit oriented at PA $-$25$^{\circ}$ (see Fig.\,3). Spectra reduction was carried out following standard procedures in 
{\sc xvista}.\footnote{{\sc xvista} was originally developed as Lick Observatory Vista. It is 
currently maintained by Jon Holtzman at New Mexico State University and is available at
http://ganymede.nmsu.edu/holtz/xvista.} In particular, we derived the median of the three 
spectra to detect and eliminate cosmic rays, and the median spectrum was used as our final 
spectrum. The seeing was $\sim$2$\farcs$5 during the observations. Spectrophotometric standards 
stars were observed on the same night as the object for flux-calibration. The final flux-calibrated
spectrum of Me\,2-1 is shown in Figure\,5.

\begin{figure}
%\vspace{302pt}
\begin{center}
\includegraphics[width=\columnwidth]{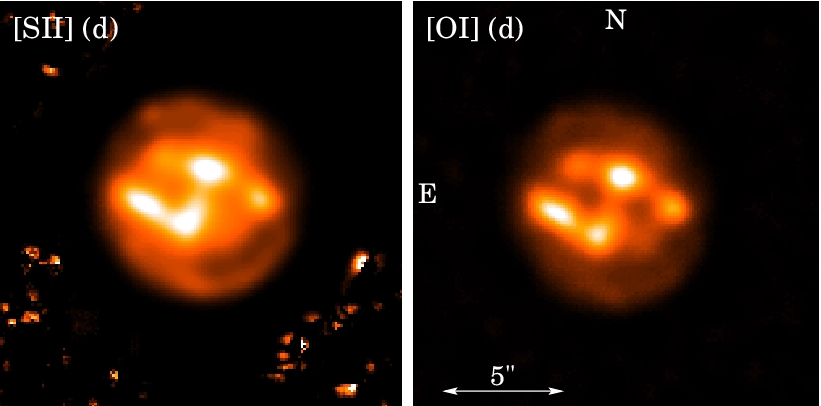}
\caption{Narrow-band [S\,{\sc ii}](d) and [O\,{\sc i}](d) of Me\,2-1. Intensity levels are 
linear. Some artifacts resulting from the deconvolution process are observed in the [S\,{\sc ii}] image. 
Orientation and spatial scale are identical in the two panels and indicated in the [O\,{\sc i}](d) image. }
\end{center}
\end{figure}

\section{Results}

\begin{figure}
%\vspace{302pt}
\begin{center}
\includegraphics[width=80mm]{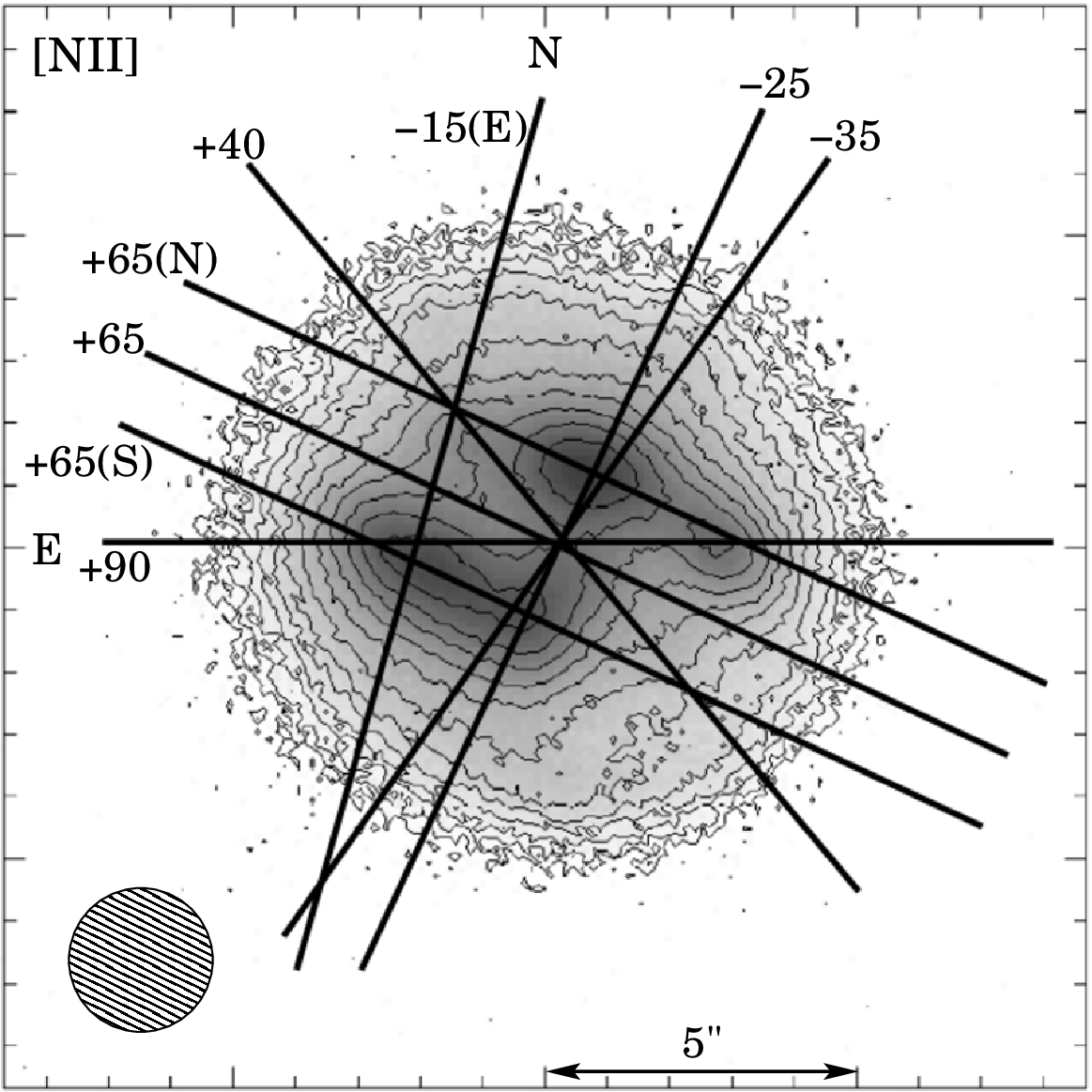}
\caption{Positions of the slits used for high-resolution spectroscopy superimposed on 
a gray scale and contours reproduction of the [N\,{\sc ii}] image of Me\,2-1. 
The slits are labeled by their PA (see the text and Table\,2 for such details as, e.g., slit width). A long-slit at PA $-$25$^{\circ}$ was also used for
the intermediate-resolution spectra (see Sect.\,2.3). The scale and orientation of the 
image are indicated. The dashed circle represents the mean seeing in the long-slit spectra (see Table\,2).
}
\end{center}
\end{figure}

\begin{table}
\centering  
\caption{Intrinsic emission line intensities in Me\,2-1 ($I$(H$\beta$)=100.0) and
derived physical parameters for the nebula.}                           
\begin{tabular}{lcc}
  \hline
  \hline
  Emission line          & $f$$_{\lambda}$ & $I$($\lambda$) \\
  \hline
\ion{He}{ii}\,4339\tablefootmark{a}       &   0.157 &     2.0$\pm$0.1 \\
H$\gamma$\,4340          &   0.157 &    44.4$\pm$0.4 \\ 
$[\ion{O}{iii}]$\,4363   &   0.149 &    21.1$\pm$0.3 \\ 
\ion{He}{i}\,4471        &   0.115 &     1.2$\pm$0.2 \\ 
\ion{He}{ii}\,4540       &   0.094 &     3.4$\pm$0.2 \\ 
\ion{N}{iii}\,4640       &   0.063 &     2.3$\pm$0.2 \\ 
\ion{He}{ii}\,4686       &   0.050 &    84.4$\pm$0.4 \\ 
$[\ion{Ar}{iv}]$\,4711   &   0.042 &     7.9$\pm$0.2 \\ 
$[\ion{Ne}{iv}]$\,4724   &   0.039 &     1.9$\pm$0.1 \\ 
$[\ion{Ar}{iv}]$\,4740   &   0.034 &     6.4$\pm$0.2 \\ 
\ion{He}{ii}\,4859\tablefootmark{a}       &   0.000 &     4.2$\pm$0.1 \\
H$\beta$\,4861           &   0.000 &   100.0$\pm$0.4 \\ 
$[\ion{O}{iii}]$\,4959   &  -0.026 &   446.2$\pm$1.4 \\ 
$[\ion{O}{iii}]$\,5007   &  -0.038 &  1320.6$\pm$4.0 \\ 
\ion{He}{ii}\,5411       &  -0.126 &     7.1$\pm$0.1 \\ 
$[\ion{Cl}{iii}]$\,5517  &  -0.145 &     0.5$\pm$0.1 \\ 
$[\ion{Cl}{iii}]$\,5537  &  -0.149 &     0.5$\pm$0.1 \\ 
$[\ion{N}{ii}]$\,5755    &  -0.185 &     0.3$\pm$0.1 \\ 
\ion{C}{iv}\,5806        &  -0.193 &     0.2$\pm$0.1 \\ 
\ion{He}{i}\,5876        &  -0.203 &     4.0$\pm$0.1 \\ 
$[\ion{O}{i}]$\,6300     &  -0.263 &     1.7$\pm$0.1 \\ 
$[\ion{S}{iii}]$\,6312   &  -0.264 &     2.1$\pm$0.1 \\ 
$[\ion{O}{i}]$\,6363     &  -0.271 &     0.6$\pm$0.1 \\ 
\ion{He}{ii}\,6407       &  -0.277 &     0.4$\pm$0.1 \\ 
$[\ion{Ar}{v}]$\,6435    &  -0.281 &     1.1$\pm$0.1 \\ 
\ion{He}{ii}\,6527       &  -0.293 &     0.5$\pm$0.1 \\ 
$[\ion{N}{ii}]$\,6548    &  -0.296 &     3.7$\pm$0.1 \\ 
\ion{He}{ii}\,6560\tablefootmark{a}       &  -0.298 &    11.4$\pm$0.2 \\
H$\alpha$\,6563          &  -0.298 &   281.5$\pm$1.2 \\ 
$[\ion{N}{ii}]$\,6583    &  -0.300 &    12.1$\pm$0.1 \\ 
\ion{He}{i}\,6678        &  -0.313 &     1.9$\pm$0.1 \\ 
$[\ion{S}{ii}]$\,6716    &  -0.318 &     2.3$\pm$0.1 \\ 
$[\ion{S}{ii}]$\,6731    &  -0.320 &     2.8$\pm$0.1 \\ 
\ion{He}{ii}\,6891       &  -0.341 &     0.5$\pm$0.1 \\ 
$[\ion{Ar}{v}]$\,7006    &  -0.356 &     2.3$\pm$0.1 \\ 
\ion{He}{i}\,7065        &  -0.364 &     1.3$\pm$0.1 \\ 
$[\ion{Ar}{iii}]$\,7136  &  -0.374 &    10.2$\pm$0.1 \\ 
\ion{He}{ii}\,7178       &  -0.379 &     1.2$\pm$0.1 \\ 
$[\ion{Ar}{iv}]$\,7236   &  -0.387 &     0.3$\pm$0.1 \\ 
$[\ion{Ar}{iv}]$\,7264   &  -0.391 &     0.4$\pm$0.1 \\ 
\ion{He}{i}\,7281        &  -0.393 &     0.3$\pm$0.1 \\ 
$[\ion{O}{ii}]$\,7320    &  -0.398 &     1.9$\pm$0.1 \\ 
$[\ion{O}{ii}]$\,7330    &  -0.400 &     1.8$\pm$0.1 \\ 
\hline
$c$(H$\beta$)            &         &    0.15$\pm$0.01 \\
log\,$F$(H$\beta$)      &   (erg\,cm$^{-2}$\,s$^{-1}$)      &  $-$11.64$\pm$0.02 \\ 
%(erg\,cm$^{-2}$\,s$^{-1}$)  &         &           \\
\hline
$T$$_{\rm e}$($[\ion{N}{ii}]$)   &    (K)     &   13880$\pm$1540 \\
$T$$_{\rm e}$($[\ion{O}{iii}]$)  &    (K)     &   13900$\pm$120 \\
$N$$_{\rm e}$($[\ion{S}{ii}]$)   &   (cm$^{-3}$)      &    1575$\pm$345 \\
$N$$_{\rm e}$($[\ion{Cl}{iii}]$) &  (cm$^{-3}$)       &    3260$\pm$2545 \\
$N$$_{\rm e}$($[\ion{Ar}{iv}]$)  &  (cm$^{-3}$)       &    1750$\pm$640 \\
\hline
\end{tabular}
\tablefoot{ 
  \tablefootmark{a}{The He\,{\sc ii} Pickering emission lines are not observed because they are blended with the corresponding Balmer line.
    The listed (predicted) intensities
    have been obtained from the theoretical intensity ratio of the Pickering lines with respect to that of \ion{He}{ii}\,4686 (see text).}
}\end{table}

\subsection{Morphology}

The narrow-band images presented in this paper, in particular those obtained with filters covering low-excitation emission lines, reveal
for the first time the structure of Me\,2-1 and its point-symmetry (Figs.\,1 and 2). We will start by describing the [N\,{\sc ii}] image. 

The [N\,{\sc ii}] image shows a slightly elliptical nebula with a minor axis at PA$\sim$$-$30$^{\circ}$ that contains two pairs of bright,
PS knots and hints of other structures (Fig.\,1). More morphological details can be recognized in the [N\,{\sc ii}](d) image (Fig.\,1), which reveals
a faint, knotty elliptical ring of $\simeq$7$\farcs$8$\times$8$\farcs$6 in diameter with the minor axis at PA $\sim$$-$30$^{\circ}$, and two elongated bright
structures, hereafter referred to as caps, that are slightly curved and mainly oriented at PA $\simeq$+65$^{\circ}$, and that extend over a range of PAs of
$\sim$140$^{\circ}$ on the plane of the sky, as measured from
the CS. The caps contain three pairs of bright knots labeled AA', BB', and CC' (Fig.\,1) that present an exceptional point-symmetry with respect
to the CS and that will be hereafter referred to as the PS knots. The inter-knot angular distance along each cap is 1$\farcs$8--2$\farcs$5. Knots CC' are noticeably
fainter than knots AA' and BB', and knot B presents an extension toward the center of the nebula. Knots BB' are oriented at PA $\simeq$$-$30$^{\circ}$, which coincides
with the ring axis. The deconvolved angular size (FWHM) of the PS knots is $\simeq$1$''$. Except for the elliptical ring, the caps, and PS knots, the rest of
the nebular [N\,{\sc ii}] emission is very faint.

\begin{figure*}
%\vspace{302pt}
\begin{center}
\includegraphics[width=170mm]{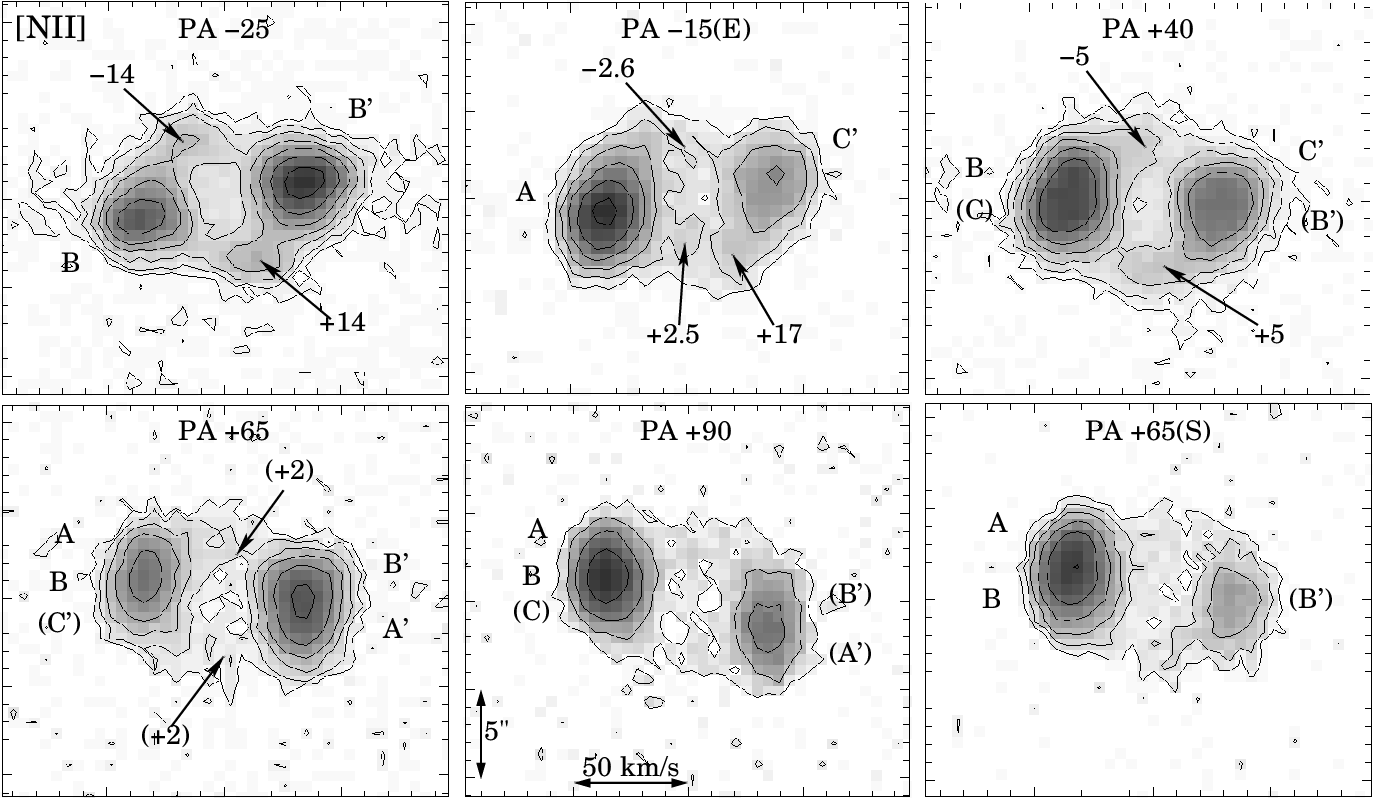}
\caption{Gray scale and contour PV maps of the [N\,{\sc ii}]$\lambda$6583 emission line. The gray levels and contours are logarithmic, the latter
  separated by a factor of two in linear intensity. The PA of each PV map is indicated and its position in the PV map marks the orientation of the slit,
  in agreement with the orientations of the slits marked in Fig\,3.
  Spatial and radial velocity scales are indicated in the PV map at PA +90$^{\circ}$. The letters indicate the knots that are mainly covered by each
slit (see Fig.\,1) and the numbers are radial velocities (km\,s$^{-1}$) with respect to the systemic velocity of the nebula (Sect.\,3.2). }
\end{center}
\end{figure*}

The morphology in the [S\,{\sc ii}](d) and [O\,{\sc i}](d) images (Fig.\,2) is similar to that in the [N\,{\sc ii}](d) one. The caps, PS knots, and parts
of the elliptical ring are identified in these images, although knot C is difficult to recognize in the [S\,{\sc ii}](d) image.
As was mentioned above, the [S\,{\sc iii}]$\lambda$6312 emission line is included in the [O\,{\sc i}] filter (Table\,1). Nevertheless, the  prominence of
the PS knots in the [N\,{\sc ii}] and [S\,{\sc ii}] images strongly suggests that the [O\,{\sc i}]$\lambda$6300 emission could be the dominant one
in the [O\,{\sc i}] filter.

The H$\alpha$ image (Fig.\,1) shows very faint extended emission up to $\simeq$7$''$ from the center.\footnote{A slight elongation along
PA $\sim$$-$40$^{\circ}$ seems to be present in the faint emission, although our data do not allow us to conclude whether this elongation is real
or corresponds to an artifact, because it is also observed in a bright star (not shown in Figs.\,1 and 2) toward the west of Me\,2-1.}
The H$\alpha$(d) image shows the elliptical ring and the caps, although individual PS knots cannot be identified.
In the [O\,{\sc iii}] image, Me\,2-1 appears to be round, with a radius
of $\simeq$5$''$, and the emission peaks at the center of the nebula. The [O\,{\sc iii}](d) image reveals
the elliptical ring, bright curved structures that corresponds to the caps, and the CS at the center of the nebula.
The deconvolved size (FWHM) of the caps in the [O\,{\sc iii}] and H$\alpha$ images is $\simeq$1$''$. In the H$\alpha$(d) and [O\,{\sc iii}](d) images the
intensity contrast between the ring and the inner nebular regions (different from the caps) is much lower than observed in the [N\,{\sc ii}](d) one,
indicating that material interior to the ring is of high excitation. 

A comparison of the images described above with the $HST$ ones (see, e.g., SPGL04\footnote{Fig.\,1 in SPGL04 should be rotated 150.5 degrees counterclockwise
to get North at the top and East to the left as in our Fig.\,1.}) shows that the barrel-like structure seen in the $HST$ images can be
identified with the caps observed in our images. The knotty elliptical ring can also be recognized in the $HST$ images. None of the {\it HST} images shows the
PS knots detected only in the low-excitation emission line images.

\subsection{Internal kinematics}

\begin{table}
 \centering  
\caption{Ionic abundances in Me\,2-1 relative to H$^+$.}                           
\begin{tabular}{lr}
  \hline
  \hline
Ion                           & Abundance  \\
\hline
He$^{+}$ ($\times$$10^{2}$) &   3.33$\pm$0.06 \\ 
He$^{2+}$ ($\times$$10^{2}$) &   6.84$\pm$0.04  \\ 
O$^{0}$ ($\times$$10^{6}$)  &   1.2$\pm$0.4  \\ 
O$^{+}$ ($\times$$10^{6}$) &   7.1$\pm$3.5  \\ 
O$^{2+}$ ($\times$$10^{4}$) &   1.7$\pm$0.4  \\ 
N$^{+}$ ($\times$$10^{6}$) &   1.1$\pm$0.2  \\ 
Ne$^{3+}$ ($\times$$10^{4}$) &   2.1$\pm$2.2  \\
S$^{+}$ ($\times$$10^{8}$) &   7.5$\pm$1.6  \\ 
S$^{2+}$ ($\times$$10^{6}$) &   1.5$\pm$0.6  \\ 
Cl$^{2+}$ ($\times$$10^{8}$) &   2.83$\pm$0.81  \\ 
Ar$^{2+}$ ($\times$$10^{7}$) &   4.72$\pm$1.06  \\ 
Ar$^{3+}$ ($\times$$10^{7}$) &   5.57$\pm$1.17  \\ 
Ar$^{4+}$ ($\times$$10^{7}$) &   2.25$\pm$0.48  \\ 
\hline
\end{tabular}
\end{table}

\begin{table}
 \centering  
\caption{Elemental abundances in Me\,2-1 derived in this paper and in SPGL04.}                           
\begin{tabular}{lcc}
  \hline
  \hline
Element                 &  Abundance      & Abundance   \\
                        & (this paper)     & (SPGL04)   \\
\hline
\hline
He/H                    & 0.102$\pm$0.001  &   0.1  \\ 
O/H ($\times$$10^{4}$)   & 4.02$\pm$0.82   &  5.1--5.3 \\ 
N/H  ($\times$$10^{5}$)  & 5.0$\pm$3.3     & 5.1--5.7 \\
Ar/H ($\times$$10^{6}$)  & 1.73$\pm$0.02   &  1.6  \\ 
S/H  ($\times$$10^{6}$)  & 3.9$\pm$1.8     & 3.5--9.1 \\ 
\hline
\end{tabular}
\end{table}

The PV maps of the emission lines mainly show a velocity ellipse (Figs.\,4, A1, A2). The radial
velocity splitting at the spatial center of the velocity ellipses amounts to 33.0$\pm$1.6, 35.2$\pm$1.4, 50.0$\pm$1.5, and 67.0$\pm$1.3 in He\,{\sc ii}, H$\alpha$,
[O\,{\sc iii}], and [N\,{\sc ii}], respectively, averaged from all observed PAs. From the radial velocity centroid of the observed emission lines,
also measured at the spatial center of the velocity ellipses, we obtain a systemic velocity with respect to the Local Standard of Rest
of +53.7$\pm$2.5\,km\,s$^{-1}$, in excellent agreement
with the value of +53.6$\pm$5.3\,km\,s$^{-1}$ obtained by Schneider et al. (1983). Throughout this paper, internal radial velocities in the nebula
are quoted with respect to the systemic velocity.

The caps and PS knots are easily recognizable in the PV maps of the [N\,{\sc ii}] emission line because of their brightness (Fig.\,4). Individual knots cannot be
distinguished in the PV maps due to their small inter-knot angular separation and inadequate seeing in the long-slit spectra to resolve them spatially.
Nevertheless, Figs.\,1, 3, and 4 allow us to guess which PS knots are contributing at each PA, as is indicated in Fig.\,4. The southeastern (SE) caps and PS knots
are blueshifted,
the northwestern (NW) ones are redshifted, and their radial velocity (in absolute value) is $\simeq$32--35\,km\,s$^{-1}$. The velocity width (FWHM) in the caps amd PS knots
is $\simeq$21--27\,km\,s$^{-1}$.

Some of the [N\,{\sc ii}] PV maps also show faint knots or intensity enhancements at or close to the spatial tips of the velocity ellipse (see Fig.\,4). Faint
knots toward the NW are blueshifted, while those toward the SE are redshifted, and their radial velocities seem to change with the PA in a systematic manner, reaching a
maximum of $\simeq$$\pm$14\,km\,s$^{-1}$ at PA $-$25$^{\circ}$, decreasing to $\simeq$$\pm$5\,km\,s$^{-1}$ at PA +40$^{\circ}$, and with a somewhat uncertain
minimum value of $\simeq$$\pm$2\,km\,s$^{-1}$ at PA +65$^{\circ}$ (see Fig.\,4).  A comparison of the PV maps and [N\,{\sc ii}](d) image indicate that
the faint knots trace emission from the elliptical ring (Fig.\,1). Consequently, the maximum radial velocity in the ring seems to be observed along its minor axis
and minimum radial velocity along its major axis. 

Other regions of the velocity ellipse are very faint and superimposed by the strong emission from the PS knots. At PAs +90$^{\circ}$ and +65$^{\circ}$(S)
a velocity ellipse is difficult to recognize, probably because of the seeing. In some PV maps it seems that the emission from the PS knots extend beyond the
velocity ellipse. Probably this effect is not real, as the images show, and rather a result of the large brightness difference between the PS knots and velocity ellipse.
This effect may also cause the velocity ellipse to seem to be tilted at some PAs (e.g., PA +90$^{\circ}$) on the PV maps, although this may not be the case.
In any case, the presence of a velocity ellipse in all of the PV maps (see also below) indicates the existence of a spheroidal shell interior to the ring.

The PV maps of the H$\alpha$ emission line (Fig.\,A1) show faint extended emission that corresponds to the faint halo or attached shell observed in the 
    image (Fig.\,1). In the central regions, a velocity ellipse of $\simeq$2$\farcs$5 in radius is observed, whose emission is dominated by two bright features
that are separated by $\sim$38\,km\,s$^{-1}$ in radial velocity and correspond to the caps (Fig\,1), with the NW and SE caps redshifted and blueshifted, respectively.
The PV maps of He\,{\sc ii} (Fig\,A1) mainly show a
velocity ellipse, although with some irregularities at some PAs. 

The PV map of the [O\,{\sc iii}] emission line (Fig\,A2) shows emission up to $\sim$6$''$ from the center, as is observed in the [O\,{\sc iii}] image (Fig.\,1), and is
dominated by two very bright emission features separated by $\simeq$48\,km\,s$^{-1}$ in radial velocity, corresponding to the caps. The velocity width (FWHM)
in the caps is $\simeq$20\,km\,s$^{-1}$, while their spatial extent is uncertain because of the seeing. The caps form a part of a velocity ellipse of
$\simeq$3$\farcs$2 in radius. At the spatial tips of this ellipse, two protrusions are observed with a radial velocity of $\sim$$\pm$5\,km\,s$^{-1}$ -- the
NW and SE protrusions are blueshifted and redshifted, respectively -- and they are related to the elliptical ring.

Finally, it is important to emphasize that the kinematics of the caps and PS knots traces very well that of the velocity ellipse in the three observed emission lines,
strongly suggesting that the spheroidal shell and caps or PS knots are morphokinematically related to each other.

\subsection{Physical conditions and chemical abundances}

\begin{figure}
%\vspace{302pt}
\begin{center}   
\includegraphics[width=\hsize]{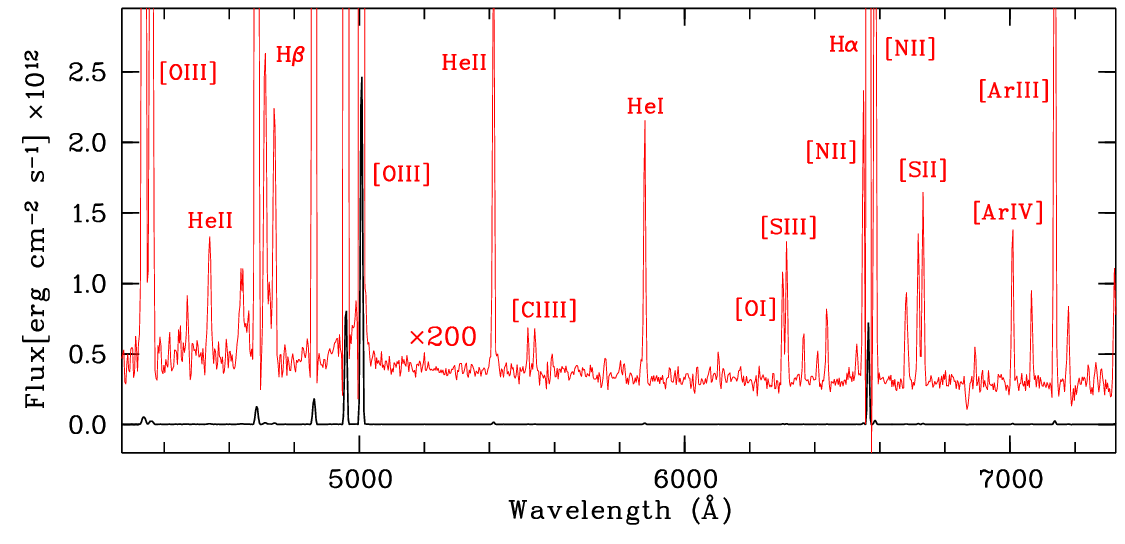}
\caption{Flux-calibrated spectrum of Me\,2-1 presented at two scales to show the faint emission lines. Some emission lines are labeled.}
\end{center}
\end{figure}

As was previously found by other authors (SPGL04 and references therein), relatively strong medium- and high-excitation emission
lines are prominent in Me\,2-1, while low-excitation emission lines tend to be weak or very weak (Fig.\,5). It should be noted that our
intermediate-resolution spectrum of Me\,2-1 is mostly dominated by emission from the caps or PS knots, in particular from knots BB' (see Fig.\,3).
This is particularly relevant for the [O\,{\sc iii}], [N\,{\sc ii}], and, probably, H$\alpha$ emission lines, for which the PV maps clearly show
that the caps or PS knots largely dominate the emission in these lines (Figs.\,4, A1, and A2). 

We obtained the logarithmic extinction coefficient, $c$(H$\beta$), from the observed Balmer emission lines. In our spectrum the
He\,{\sc ii}$\lambda$4686/H$\beta$ observed flux ratio is relatively high, $\simeq$0.83. Therefore, we corrected the observed flux of 
the Balmer emission lines, taking into account the flux of the corresponding
Pickering He\,{\sc ii} emission lines, which was obtained from the theoretical ratios between the Pickering and the He\,{\sc ii}$\lambda$4686 emission lines.
Nevertheless, the contribution of the Pickering lines to the Balmer ones is small, amounting to between $\simeq$4 and $\simeq$8\%. We obtain $c$(H$\beta$) $\simeq$0.15,
which is smaller than the values between $\simeq$0.25 and $\simeq$0.34 usually found, although similar to the value of $\simeq$0.11 obtained by Kaler et al.
(1987). With our value of $c$(H$\beta$) and the extinction law by Cardelli et al. (1989) we obtained the intrinsic emission line intensities that are listed
in Table\,3 (including the predicted ones for the Pickering emission lines), together with $c$(H$\beta$) and the observed H$\beta$ flux.

From Table\,3, we obtain [N\,{\sc ii}]6583/H$\alpha$, [O\,{\sc iii}]5007/H$\alpha$, and [N\,{\sc ii}]6583/[O\,{\sc iii}]5007 line intensity ratios of
$\simeq$0.04, $\simeq$4.67, and $\simeq$0.01, respectively, which should mainly reflect the excitation of the caps or PS knots. It is obvious that these structures
are of high excitation and would probably have gone unnoticed in images obtained with filters (e.g., H$\alpha$+[N\,{\sc ii}] or R) that did not cover
exclusively low-excitation emission lines. 

We analyzed the spectrum with {\sc anneb} (see Olgu\'{\i}n et al. 2011), which uses {\sc iraf}\,2.16, to obtain $T$$_{\rm e}$, $N$$_{\rm e}$, and the
ionic abundances. Table\,3 also lists the values of $T$$_{\rm e}$ and $N$$_{\rm e}$, derived from different emission line ratios, and Table\,4 lists the ionic abundances.
To further check our results, we also used the code {\sc pyneb} (v.1.1.14; Luridiana et al. 2015) and, as has already been reported for other PNe (e.g.,
Aller et al. 2021), we also found that both codes give very similar results. In general, the derived physical parameters and ionic abundances are comparable
to or compatible with previous ones (Aller et al. 1981; Kaler 1985; Moreno et al. 1994; SPGL04). 

Table\,5 lists the elemental abundances obtained with the ionization correction factors of Delgado-Inglada et al. (2014). We also include
in Table\,5 the abundances determined by SPGL04 for comparison purposes. In general, there is a reasonable agreement between the two sets of
abundances, although we note the large error in our nitrogen abundance due to the faintness of the [N\,{\sc ii}] emission lines. In any case,
our abundances are compatible with a low-mass progenitor of Me\,2-1, as was already concluded by SPGL04.

\section{Discussion}

The data presented above have revealed that Me\,2-1 consists of an elliptical ring, a spheroidal shell interior to the ring, and two bright curved caps
containing three pairs of PS knots that are only observed in the images of low-excitation emission lines. These structures are surrounded by a faint halo or
attached shell. In the following, we attempt to reconstruct the 3D structure of Me\,2-1 (Sect.\,4.1), comment on the
properties and nature of the caps or PS knots, and on the unusual 3D structure of Me\,2-1 (Sect.\,4.2), and discuss the formation of this PN (Sect.\,4.3).

\subsection{Reconstruction of the three-dimensional structure}

\begin{figure*}
%\vspace{302pt}
  %\begin{center}
  \centering
  \vspace{10pt}
  \includegraphics[width=17cm]{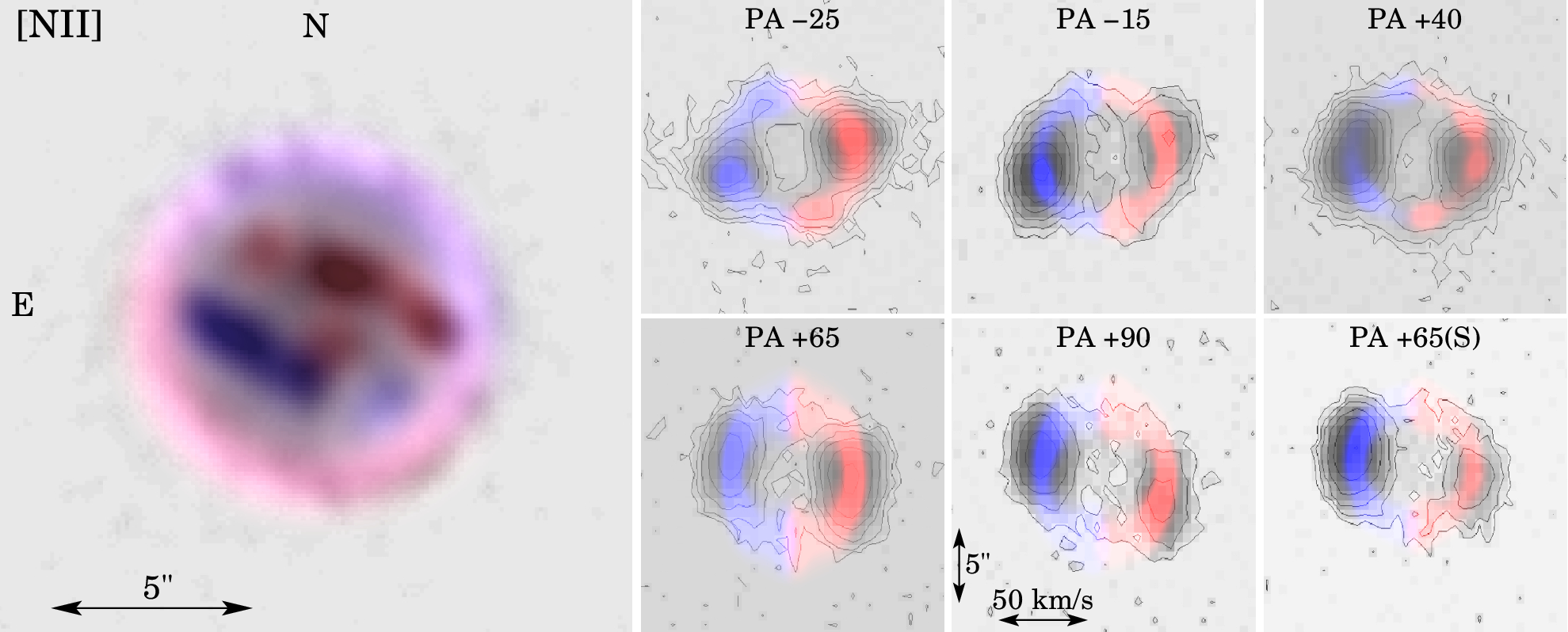}
  \includegraphics[bb=17 12 675 190,width=17cm,clip]{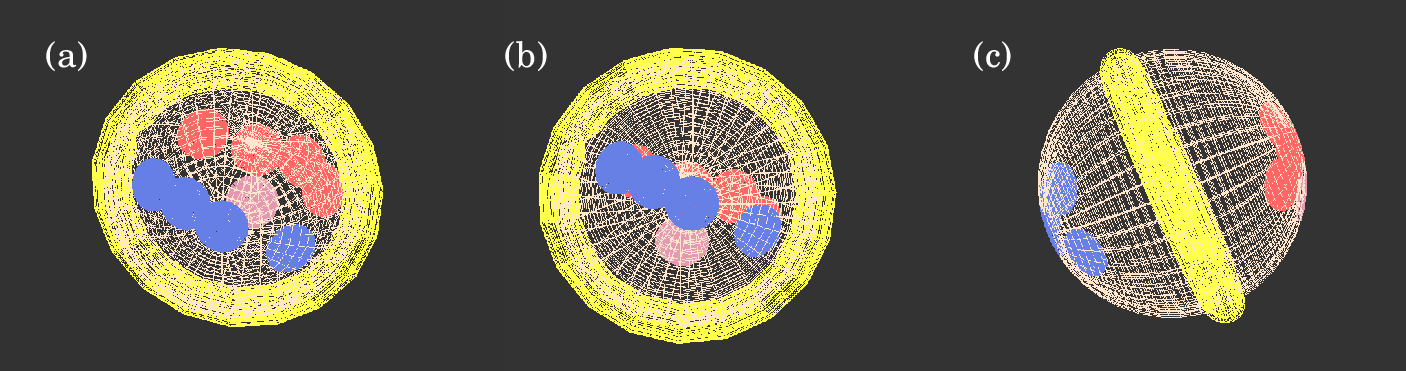}
  \caption{({\it top}) Deconvolved image and PV maps of the [N\,{\sc ii}] emission line with the {\sc shape} reconstruction superimposd on top.
    Blue and red colors are used for regions that are blueshifted and redshifted, respectively, with respect to the systemic velocity. Otherwise, markings
    are the same as in Fig.\,4 for the PV maps. ({\it Bottom}) 3D views of our {\sc shape} reconstruction of Me\,2-1: (a) from the observer; (b) pole-on;
    (c) side view. Yellow represents the equatorial ring, white the spheroidal shell, blue and red circles the caps or knots, and light red the knot at
    the center of the nebula (see text).}
%\end{center}
\end{figure*}

To reconstruct the 3D structure of Me\,2-1, we used the tool {\sc shape} (Steffen et al. 2011) and the [N\,{\sc ii}](d) image and [N\,{\sc ii}] PV maps.
Before starting with {\sc shape}, it is worth using the data to obtain estimates for some
morphokinematical parameters that may constrain the initial ones in {\sc shape}. Clues about some parameters are provided by the morphokinematical
properties of the elliptical ring; namely, that the maximum or minimum radial velocity seems to be observed along its minor or major axis. This result is
compatible with a circular ring whose axis is oriented at PA $\simeq$$-$30$^{\circ}$, tilted $\simeq$24$^{\circ}$ with respect to the observer, and that expands
at $\simeq$34\,km\,s$^{-1}$. Knots BB' are observed along  the axis of the ring, suggesting that they may be polar knots. If so, their velocity
is $\simeq$35\,km\,s$^{-1}$. As was already mentioned, the PV maps (Fig.\,4) indicate that the caps and PS knots and the spheroidal shell share a common
morphokinematics. Therefore,
we assumed that the caps and PS knots are located on the surface of the spheroidal shell. For our reconstruction, we considered the caps and PS knots as circular
patches on the surface of the spheroidal shell to simulate the length of the caps and did not consider the individual PS knots that are not spatially resolved
in the PV maps. The similarity between the guessed equatorial and polar velocity further suggests that the spheroidal shell does not present a
large axis ratio. In the reconstruction we assumed homologous expansion ($V$[km\,s$^{-1}$]=K$\times$$r$[arcsec]), and two velocity laws: one for the ring
  and other for the shell and caps or PS knots. Finally, an important point to be taken into account in {\sc shape} is the relatively poor seeing in our
spectra as compared with that
in the (deconvolved) images. We tried several ways to simulate the seeing in {\sc shape} and the best one we found was to consider a slit width similar to the
seeing (Table\,2); in this way we were able to recover in the synthetic PV maps the emission from the two caps or PS knots at the different slit PAs,
including, for example, those at PAs +65$^{\circ}$ and +65$^{\circ}$(E). 

Under these assumptions, our best 3D reconstruction of Me\,2-1 is shown in Fig.\,6 superimposed on the [N\,{\sc ii}](d) image and [N\,{\sc ii}] PV maps. The
resulting structural components and their morphokinematical properties are: (1) a ring of $\simeq$4$\farcs$3 in radius with its axis
oriented at PA $\simeq$$-$28$^{\circ}$ and tilted by $\simeq$24$^{\circ}$ with respect to the observer, and with an expansion velocity of $\simeq$31\,km\,s$^{-1}$
($V$[km\,s$^{-1}$]=7.56$\times$$r$[arcsec]); (2) a spherical shell of $\simeq$4$\farcs$0 in radius, expanding at $\simeq$35\,km\,s$^{-1}$
($V$[km\,s$^{-1}$]=8.57$\times$$r$[arcsec]); and (3) the caps and PS knots located on the surface of the spherical shell (hence, with the same velocity law as the shell), also
expanding at $\simeq$35\,km\,s$^{-1}$ and covering an angle of $\simeq$75$^{\circ}$, seen from the CS. There seem to be slight differences between the sizes and expansion
velocities of the ring and shell but they are within the errors of $\pm$0$\farcs$2 for sizes and $\pm$2\,km\,s$^{-1}$ for expansion velocities.
Finally, the extension of knot\,B toward the center of the nebula seems to be associated with the NW cap. Fig.\,6 also shows different views of Me\,2-1 based on our
reconstruction.

Although our 3D reconstruction reproduces the data reasonably well, some aspects should be mentioned. We have considered a spherical shell but, in principle, 
an ellipsoidal one cannot be ruled out. However, for ellipsoidal shells with an axis ratio $\geq$1.2 and a major axis perpendicular to the plane of the ring, the
velocity ellipse in the PV maps should present a tilt in the same direction as the ring at PAs around $-$28$^{\circ}$, which is not observed in the PV maps (Figs.\,4 and A2).
In addition, such an ellipsoidal shell would produce in the images, at least at low intensity levels, an elliptical nebula with the major axis at a PA of
$\sim$$-$28$^{\circ}$, which is not observed (Fig.\,1). Some regions in the PV maps at PAs, for example, $-$25$^{\circ}$ and +90$^{\circ}$, are not ``exactly'' reproduced, but this may
be due to local inhomogeneities in the velocity field and/or density distribution in the shell, as is observed in, for example, the
``spherical'' PN NGC\,7094 (Rauch 1999). In any case, the differences between reconstruction and observations are small and acceptable with the data we have analyzed.
To refine our model, images at higher resolution and long-slit spectra at higher spectral and spatial resolutions than ours will be necessary.

The morphokinematical parameters allow us to obtain the kinematical age of the structures if the distance (D) is known. Unfortunately, a Gaia distance is not available
  for Me\,2-1, and published distances cover a wide range, from 2 to 8.8\,kpc, derived from approximated or statistical methods (Moreno et al. 1994;
Cazzeta \& Maciel 2000; SPGL04; Tajitsu \& Tamura 2008 and reference therein). From the 3D reconstruction, we obtain an age of 640$\times$D[kpc]\,yr
for the ring and 560$\times$D[kpc]\,yr for the shell and caps and PS knots, with an error of $\pm$100$\times$D[kpc]\,yr. In H$\alpha$ and [O\,{\sc iii}], we also assume
a spherical shell with caps attached to it, and obtain ages of 600$\times$D[kpc]\,yr in H$\alpha$ and 580$\times$D[kpc]\,yr in [O\,{\sc iii}] for both
structures, with an error of $\pm$150$\times$D[kpc]\,yr. All kinematical ages agree within the errors. Nevertheless, as we will discuss below (Sect.\,4.2), the coincidence
in kinematical age does not necessarily imply that all the structures are due to simultaneous processes.

\subsection{The nature and properties of the caps and point-symmetric knots, and the peculiarities of Me\,2-1}

A remarkable characteristic of Me\,2-1 is the PS structures (caps and PS knots) attached to the shell. In this respect, caps and PS knots in Me\,2-1
  present extraordinary similarities to the caps in NGC\,6543 (components DD' in Miranda \& Solf 1992; see also Balick \& Hajian 2004). They expand at
  $\simeq$28--38\,km\,s$^{-1}$ in NGC\,6543 and $\simeq$35\,km\,s$^{-1}$ in Me\,2-1. In both cases they subtend an angle of $\simeq$75$^{\circ}$--80$^{\circ}$ as seen from
  the CS; evidence exists in both cases for a contribution from shock excitation although the main excitation mechanism is photoionization (see also Mari et al. 2023). Other
  similar PNe with PS knots or arcs attached to the main shell include, for example, NGC\,6309 (V\'azquez et al. 2008), KjPn\,8 (L\'opez et al. 1995, 1997),
  and NGC\,6572 (Miranda et al. 1999; Akras \&  Gon\c calves 2016). These structures are better interpreted as a result of the interaction of collimated bipolar outflows
  with the main shell if, due to particular initial physical and kinematic conditions, the collimated outflows are slowed down in the interaction and, finally, both
  structures share their morphokinematic properties (see Miranda \& Solf 1992). If the collimation axis changes its orientation with time, extended arcs or knots at
  different orientations will be observed on the shell.

  From the comments above, the caps and PS knots in Me\,2-1 most probably represent the sites where collimated outflows collide
  with the shell and are slowed down, remaining attached to the main shell. Moreover, the narrow-band and {\it HST} images show the caps (or barrel-like structure)
  as ``continuous'' extended arcs, pointing to a ``continuous'' bipolar outflow whose axis has changed its orientation by $\simeq$75$^{\circ}$ during the ejection process.
  The PS knots appear as compact structures within the caps and
  could trace episodes of enhanced mass and/or velocity ejection within more continuous outflows. It is important to note that the initial properties (e.g., velocity) of
  the outflows are lost in the interaction process and nothing can be said about when they have taken place and the time span over which they
  occurred. Nevertheless, it is reasonable to conclude that the age of the shell, caps, and PS knots derived above ($\simeq$580$\times$D[kpc]\,yr in average) represents
  the age of the shell and is an upper limit on the age of the outflows.

Unlike most PS structures in PNe, which are prominent in low-excitation emission lines (e.g., Gon\c calves et al. 2001), those in Me\,2-1 are
high-excitation structures (Sect.\,3.3), as is observed in a few PNe (e.g., NGC\,6309, V\'azquez et al. 2008). Furthermore, the expansion velocity of the shell
and caps or PS knots is noticeably higher in [N\,{\sc ii}] ($\simeq$35\,km\,s$^{-1}$) than in [O\,{\sc iii}] ($\simeq$25\,km\,s$^{-1}$) and  H$\alpha$
($\simeq$19\,km\,s$^{-1}$), following the well-known Wilson effect (Wilson 1950), which indicates that the basic excitation mechanism of all the structures is
photoionization. Nevertheless, the noticeable enhancement of the [N\,{\sc ii}], [S\,{\sc ii}], and, probably, [O\,{\sc i}] emissions in the PS knots, as compared
with those emissions in the rest of the nebula, strongly suggests that shocks might be contributing to their excitation as well. In fact, by using the
diagrams of Mari et al. (2023, their Fig.\,6), we found that several emission line ratios (e.g., [N\,{\sc ii}]/H$\alpha$, [O\,{\sc iii}]/H$\beta$,
  [N\,{\sc ii}]/[S\,{\sc ii}], [N\,{\sc ii}]5755/[N\,{\sc ii}]6583, and [O\,{\sc iii}]4363/[O\,{\sc iii}]5007, Table\,3) are located in regions where photoionization
  and (fast) shocks coexist, favoring the scenario of high-velocity outflows--shell interaction.

The combination of PS structures due to collimated outflows, and a spherical shell is unusual in PNe, as is the fact that a ring
surrounds a spherical shell. We have checked for PNe in the literature for which the physical structure has been reconstructed, and found none that resembles Me\,2-1.

\subsection{The formation of Me\,2-1: A possible scenario}

It is widely accepted that PNe with collimated outflows, multiple structures, and rings or toroids are related to interacting binary or multiple stellar
systems, those that evolve in a common envelope (CE) being particularly interesting (Boffin \& Jones 2019, and references therein).
As was already mentioned, Jacoby et al. (2021) found that the light curve of the CS of Me\,2-1 presents a periodic $\simeq$22\,d variability that these authors attribute to
irradiation of a close companion. Jacoby et al. also mention that further confirmation of the companion is necessary, given the small amplitude of
the light curve ($\simeq$0.1\%). It is worth noting that if a close binary exists in Me\,2-1, its orbital plane is expected to coincide with that of the ring
(Hillwing et al. 2016). According to our 3D reconstruction, the orbital plane should be observed at a very low
inclination of $\simeq$24$^{\circ}$. Therefore, if the detection is real, it is not surprising that the amplitude of the light curve is relatively small.
In any case, the doubts about the existence of the companion do not advise the use of
this result as a (strong) constraint to interpret the formation of Me\,2-1 (but see below). 

Regardless of whether the detection of the companion is real, the involvement of collimated outflows in the formation of Me\,2-1 and the presence of a ring
clearly point to its formation being related to a binary or multiple CSs. Numerical simulations of common envelope evolution (CEE) produce bipolar PNe for
a wide range of masses
(say, 0.2--0.7\,M\,$_{\sun}$) of the stellar companion (e.g., Iaconi \& De Marco 2019; Reichardt et al. 2019; Garc\'{\i}a-Segura et al. 2018, 2022; Zou et al. 2022;
Ondratschek et al. 2022; and references therein), which is incompatible with the virtually spherical shell of Me\,2-1. An alternative to CEE could be grazing envelope
evolution (Soker 2015) but it produces highly nonspherical PNe rather than spherical ones (Soker 2020).

Within the CEE scenario, the formation of a spherical PN seems to require the companion to have a planetary mass. Rapoport et al. (2021) analyzed the influence of
planets with 1--10\,M$_{\rm J}$ (M$_{\rm J}$ = Jupiter mass) that enter a CEE on the final shape of PNe. They found that the resulting PN may present a low degree of asphericity
for planets with more than about one Jupiter mass. Moreover, following Rapoport et al. (2021), nonspherical morphologies may also be expected for substellar
companions such as, for example, brown dwarfs, that enter a CEE. The spherical shell of Me\,2-1 would then be compatible with a planet of $\la$1\,M$_{\rm J}$.
In this scenario, however,
  the caps or PS knots are difficult to explain because it is not clear whether planets are able to generate collimated outflows. An indirect possibility is that the planet
  is tidally disrupted, forming an accretion disk around the CS from which collimated outflows are ejected (Soker 1996, 2020; Nordhaus \& Blackman 2006;
  Guidarelli et al. 2019). As is mentioned by Guidarelli et al. (2019), the resulting accretion disk--CS system behaves similarly to those found in many other kind of objects
  and can generate high-velocity collimated outflows. Moreover, these disk-driven jets can persist until the PNe phase (Blackman et al. 2001; Norhaus \& Blackman 2006).
  This scenario may explain the polar jets or ansae observed in elliptical PNe such as, for example, NGC\,3242, NGC\,6826, and NGC\,7354 (Soker 1996) and also
  the collimated outflows 
  in Me\,2-1. However, it cannot account for the large angle over which caps and PS knots extend, which requires a ``well defined'' precession or wobbling of the accretion disk.

  So far, we have discussed above CEE scenarios assuming a single companion. However, this situation may not be realistic. Observations of NGC\,3132 obtained with $JWST$
  require a system of at least four stars with different masses to explain the formation of the nebula (De Marco et al. 2022). Triple stellar systems have
been detected or proposed in several PNe such as, for example, NGC\,246, Lotr\,5, Sp\,3, Sh\,2-71, and NGC\,6720 (Jasniewicz et al. 1987; Adam \& Mugrauer 2014; Aller et al. 2018;
Miszalski et al. 2019; Jones et al. 2019; Wesson et al. 2024). At the opposite extreme of stellar evolution, many main-sequence stars are known to be accompanied by
planetary systems. Some of these planets can survive until the asymptotic giant branch (AGB) phase of the progenitor and be engulfed. Jupiter-like planets or candidates
and planetesimals
  have been detected in white dwarfs (Manser et al. 2019; Vanderburg et al. 2020; Blackman et al. 2021). Debris disks resulting from disruption of a planet or planetary
  system exist both in white dwarfs and CSs of PNe (e.g., Manser et al. 2019; Putirka \& Xu 2021, and references therein; Marshall et al. 2023;
  Swan et al. 2024). Blackman et al.
  (2021) estimate that more than 50\% of white dwarfs contain Jupiter-like planets, a number that may be similar for central stars of PNe
  because they are the immediate progenitors of white dwarfs. 

As is noted throughout this paper, the parameters of the CS of Me\,2-1 are poorly determined and many of its properties remain unknown.
  Despite this, the physical structure of the nebula strongly suggests that some particular characteristics should be present in this CS for it to be able to
  generate such an unusual combination of components.

Based on the detection of planets and debris disks in white dwarfs and CSs of PNe, and on the nebular structure of Me\,2-1, we speculate that its CS was orbited by,
at least, two planets each with a mass of $\la$1\,M$_{\rm J}$. Both enter a CEE in the AGB phase of the progenitor: one planet is tidally disrupted, forming an accretion
disk around the CS; the other survives. Because of their small mass, the two planets do not cause particular departures from the sphericity in the CE, and the ejected
shell (the resulting PN) is virtually spherical. Additionally, some mass should be ejected or deflected toward the orbital plane to form the ring, perhaps shortly
before the start of CEE. If the shell is ejected  sometime before one of the planets is disrupted, the high-velocity collimated outflows expected from the accretion
  disk--CS system will collide with the shell and, under specific initial (and unknown) conditions, slow down and remain attached to it. In addition, the accretion
  disk might wobble or precess due to the influence of the surviving planet, resulting in the extended caps or PS knots. Numerical simulations are necessary to check whether
  this scenario is feasible.

Finally, a clue to a possible planetary companion in Me\,2-1 could be hidden in the very
  small amplitude of the light curve. Because of the very small size of a planet (compared to that of any star), the contribution
  of its observed irradiated surface to the CS brightness will be extremely small, as will the variation in that contribution along the orbit, resulting in
  a very small periodic modulation of the CS brightness. A very small amplitude would be even more favored by the combination of a planet and a small orbital inclination.
  Interestingly, the light curve amplitude in Me\,2-1 of $\simeq$0.1\% is comparable to that of $\simeq$0.15\% in the Helix nebula, where the presence
  of a planet could be possible (Aller et al. 2020). Confirming the light curve in Me\,2-1  would be crucial to investigate the possible
  existence of a planetary companion. In addition, new and more precise observations of the CS are necessary to better know its properties and constrain
  its parameters.

\section{Conclusions}

We have presented a study of Me\,2-1 based on narrow-band images obtained in several emission line filters, and high- and intermediate-resolution
long-slit spectra. The main conclusions of this paper can be summarized as follows.

The images show a slightly elliptical or round shell and two pairs of PS knots. By applying deconvolution techniques to the images, they
reveal that Me\,2-1 consists of an elliptical ring, a shell interior to the ring,
and two elongated structures (caps) observed in all images; the caps contain three pairs of bright knots (PS knots) that are exclusively observed in
the low-excitation line images and that present an exceptional point-symmetry with respect to the CS. These structures are surrounded by a faint halo or
attached shell. The observed  structures can be identified in the high-resolution long-slit spectra by their morphokinematic and brightness properties.

The reconstruction of the (3D) physical structure indicates that Me\,2-1 consists of an equatorial ring seen almost pole-on, a virtually spherical shell
interior to the ring, and the caps and PS knots located on the surface of the spherical shell. The expansion velocity of these structures is a narrow range
of $\sim$31--35\,km\,s$^{-1}$. 

The properties of the caps and PS knots strongly suggest that they represent the sites where collimated outflows, ejected along a
precessing or wobbling axis, collide with the shell and slow down, remaining attached to it. The PS knots could trace episodes of enhanced mass and/or
velocity ejection within a more continuous outflow represented by the extended caps. The combination of collimated outflows and a ring with a spherical shell
is unusual in PNe. 

The kinematics of the different emission lines indicates that Me\,2-1 is mainly photoionized, although a contribution of shock excitation in the PS knots
is indicated by their relatively high brightness in the low-excitation emission lines and their emission line ratios.

We speculate that two planets each with a mass of $\la$1\,M$_{\rm J}$, orbiting the CS of Me\,2-1 and engulfed during the AGB
phase of the CS, could account for the observed structures if one of the planets is tidally disrupted, forming an accretion disk around the CS, and the
other survives. The small mass of the planets does not produce significant deviations from the sphericity in the CE and the resulting PN is virtually spherical,
although some material of the envelope should be ejected or deflected toward the orbital plane to form the ring. The caps and PS knots would be caused by the impact against
the shell of high-velocity collimated outflows ejected from the accretion disk--CS system with the surviving planet causing precession or wobbling of that disk.

The physical parameters and chemical abundances derived from our intermediate-resolution optical spectrum are similar to those previously
obtained from spectra covering the ultraviolet to near-infrared range. The chemical abundances also indicate a relatively
low mass for the progenitor star of Me\,2-1.  
 
\begin{acknowledgements}
  We thank the staffs of NOT-ORM and OAN-SPM, in particular, to Mr. Gustavo Melgoza and Mr.
  Felipe Montalvo, for assistance during the observations, and to Dr. Eloy Rodr\'{\i}guez and Dr. Matilde Fern\'andez for useful discussions
  about light curves. We are grateful to the IAC for the use of the H$\alpha$ and [N\,{\sc ii}] narrow-band filters. LFM acknowledges support from
grants PID2020-114461GB-I00 and CEX2021-001131-S, funded by MCIN/AEI/10.13039/501100011033, and from the Consejer\'{\i}a
de Transformaci\'on Econ\'omica, Industria, Conocimiento y Universidades of the Junta de Andaluc\'{\i}a, and the European Regional Development Fund from
the European Union through the grant P20-00880. RV, LO, PFG acknowledge support from grant UNAM-PAPIIT IN106720. Based upon observations carried out
at the Observatorio Astron\'omico Nacional on the Sierra San Pedro M\'artir (OAN-SPM), Baja Califormia, M\'exico. Based on
observations made with the Nordic Optical Telescope, owned in collaboration by the University of Turku and Aarhus University,
and operated jointly by Aarhus University, the University of Turku and the University of Oslo, representing Denmark, Finland and
Norway, the University of Iceland and Stockholm University at the Observatorio del Roque de
los Muchachos, La Palma, Spain, of the Instituto de Astrof\'{\i}sica de Canarias. LFM would like to dedicate this paper to
the memory of his thesis advisor, Prof. Dr. Josef Solf, who passed away on 2023 December 31.
\end{acknowledgements}


\newpage

\begin{appendix} %First appendix

  \section{Additional figures}

  We present here PV maps of the H$\alpha$ and He\,{\sc ii} emission lines at PAs $-$25$^{\circ}$ and $+$65$^{\circ}$ in Fig.\,A1, and the PV map
  of the [O\,{\sc iii}] emission line at PA $-$35$^{\circ}$ in Fig.\,A2.
  %and the [O\,{\sc iii}](d) and F457M $HST$ images of Me\,2-1 in Fig.\,A3 for comparison purposes.

\begin{figure}
%\vspace{302pt}
\begin{center}
\includegraphics[width=8.0cm]{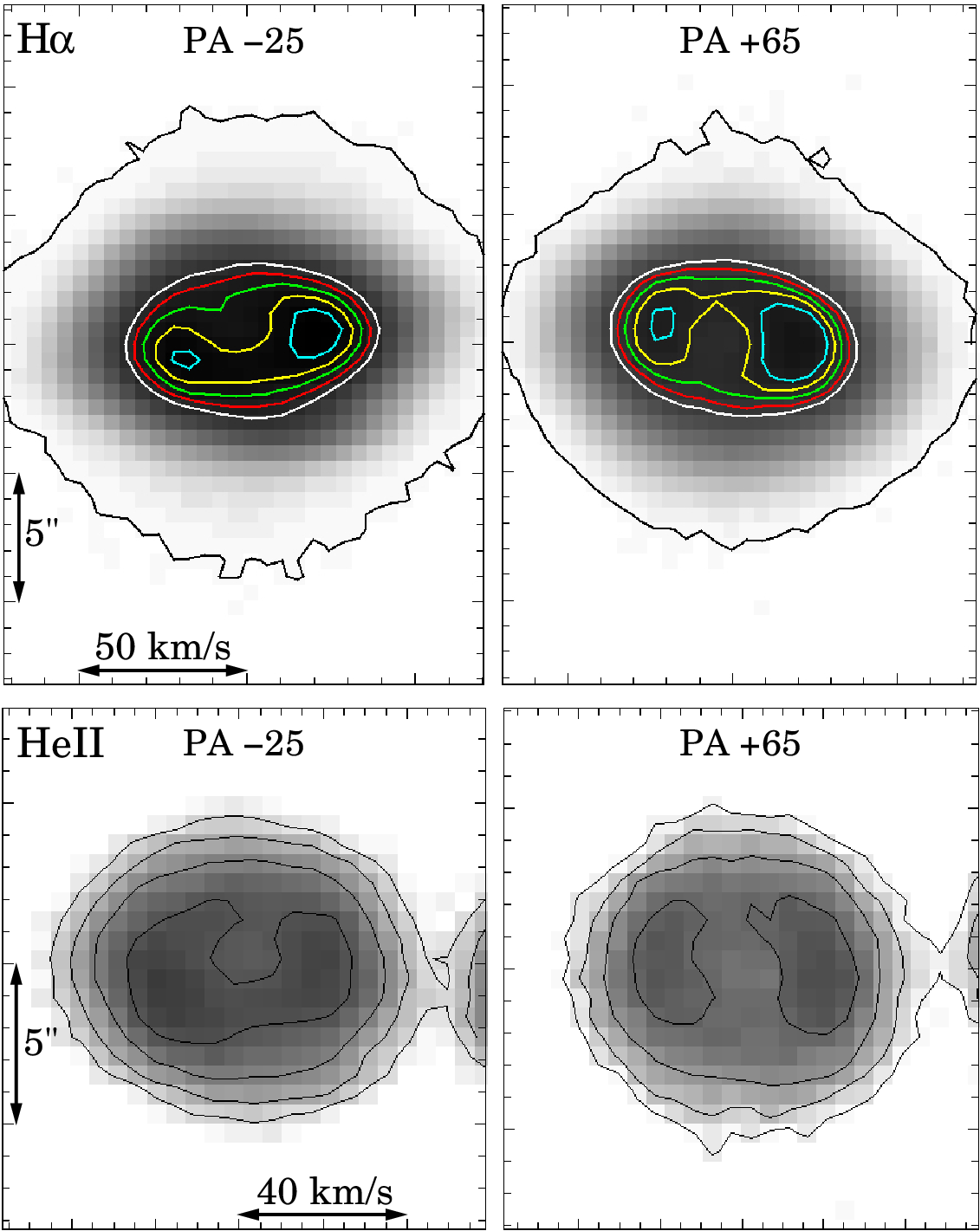}
\caption{({\it Top}) Gray scale and contour PV maps of the H$\alpha$ emission line at PAs $-$25$^{\circ}$ and $+$65$^{\circ}$.
  The contours represent increasing intensity following the color code: black, white, red, green, yellow, and cyan. The black contour is
  at $\sim$5$\sigma$ level from the background; the rest are arbitrary and have been chosen to show the spatio-kinematical structure of
  the brightest regions. ({\it Bottom}) Gray scale and contour PV maps of the He\,{\sc ii}$\lambda$6560 emission line at PAs $-$25$^{\circ}$
  and $+$65$^{\circ}$. The contours are logarithmic and separated by a factor of two in linear intensity. The emission feature toward the right
  in the two panels is the H$\alpha$ emission line. Otherwise, the markings are the same as in Fig.\,4. }
\end{center}
\end{figure}

\begin{figure}
%\vspace{302pt}
\begin{center}
\includegraphics[width=8.0cm]{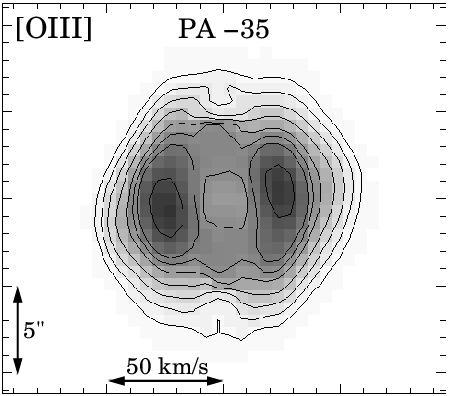}
\caption{Gray scale and contour PV map of the [O\,{\sc iii}]$\lambda$5007 emission line at PA $-$35$^{\circ}$. The contours are separated by a
  factor of two in linear intensity. Otherwise, the markings are the same as in Fig.\,4.}
\end{center}
\end{figure}

%\begin{figure}
%%\vspace{302pt}
%\begin{center}
%\includegraphics[width=\columnwidth]{me21_hst_2.eps}
%\caption{Grey-scale [O\,{\sc iii}](d) and $HST$ F547M images of Me\,2-1. The grey levels are lineal. The orientation and spatial scale are
%%  the same in both panels and indicated in the [O\,{\sc iii}](d) image.}
%\%end{center}
%\end{figure} 

\end{appendix}

\end{document}